\documentclass[12pt]{article}
\usepackage[dvips]{graphics}

\begin{document}
\begin{center}
{\Large \bfseries The spin of the $\mu$ mesons}

\vspace{.7cm}
%\bigskip
\large{E.L. Koschmieder}
\medskip

\small{Center for Statistical Mechanics, The University of Texas at 
Austin\\Austin, TX 78712 USA\\e-mail: koschmieder@mail.utexas.edu}
\smallskip
\end{center}

\bigskip
\noindent
\small {We can determine the intrinsic angular momentum of the $\mu$ 
mesons from the sum of the angular momentum vectors of the lattice 
oscillations and the sum of the spin 
vectors of the neutrinos in the lattice and the spin vector of the electric charge which the $\mu$ mesons carry. We used this neutrino lattice before to calculate the rest mass of the $\mu$ mesons. Here we learn how the apparent discrepancy between the concept of a point particle  and the finite size of a neutrino lattice is resolved. We also learn that the spin of the $\mu$ mesons 
originates exclusively from the spin of the electric charge of the 
$\mu^\pm$ mesons.}

\normalsize

\section{Introduction}
We have shown earlier [1] that the spin of the ``stable" mesons and baryons  
can be 
explained with the standing wave model [2]. In [3] we have 
determined the rest mass of the $\mu$ mesons with the concepts used in the 
standing wave model and explained why m($\mu^\pm$) is $\cong$\, 3/4
$\cdot\,
\mathrm{m}(\pi^\pm)$. According to [3] the $\mu^\pm$ mesons consist of a lattice of 
 $\nu_\mu$ (respectively $\bar{\nu}_\mu$), $\nu_e$ and 
$\bar{\nu}_e$ neutrinos which remain from the cubic neutrino lattice of the 
$\pi^\pm$ mesons after their decay, plus an electric charge. It has been argued that our explanation of the mass of the $\mu$ mesons cannot be correct because in our model the $\mu$ meson lattice has a diameter of 0.88$\cdot$10$^{-13}$ cm, whereas the $\mu$ mesons are commonly said to be point particles. However, since in our model the $\mu$ mesons consist of neutrinos, but for the electric charge, and since neutrinos do not, in a first approximation, interact with electric charge or with mass, it will not be possible to establish the size of the $\mu$ meson lattice, i.e. of the $\mu$ mesons, through conventional scattering experiments. Therefore the $\mu$ mesons will \emph{appear} to be point particles.  We will now show that the neutrino lattice does not only determine the mass of the $\mu^\pm$ mesons but 
also provides an explanation of the spin of the $\mu$ mesons which has, so 
far, not been explained.

\section{The spin of the $\mu^\pm$ mesons}

The spin s = 1/2 or the intrinsic angular momentum $\hbar$/2 of the 
$\mu^\pm$ mesons can, theoretically, be the sum of the angular momentum 
vectors of all neutrino lattice oscillations of frequency $\nu_i$ in the 
$\mu$ mesons, plus the 
sum of all spin vectors of the  $n_i$ neutrinos in the lattice, plus the 
spin 
vector of the electric charge which each $\mu$ meson carries. In a formula

\begin{equation}  j(\mu^\pm) = \sum_{i} \,j(\nu_i) + \sum_{i}j(n_i) + 
j(e^\pm) \quad   0\,\leq\, i\, \leq\,N_\mu ,  
    \end{equation}
where $N_\mu$ is the number of all neutrinos in the $\mu$ meson lattice, 
$N_\mu = 2.14\cdot 10^9$. This procedure is completely analogous to the way how the 
spin of the $\pi^\pm$ mesons is determined, only that then a cubic 
neutrino lattice is considered consisting of $\nu_\mu, \bar{\nu}_\mu, 
\nu_e$ and $\bar{\nu}_e$ neutrinos with N = 2.854$\cdot 10^9$ neutrinos.

   The lattice oscillations in the $\mu$ mesons are longitudinal and hence 
do not have an angular momentum because for each oscillation $\vec{r} 
\times \vec{p} = 0$. So the lattice oscillations do not contribute to the 
intrinsic angular momentum of the $\mu$ mesons, or $\sum_{i} j(\nu_i) = 
0$. Each of the O($10^9$) neutrinos in the $\mu$ meson lattice has, 
however, an 
angular momentum $\hbar$/2. The sum of the spin vectors of all neutrinos 
in the lattice of the $\mu$ mesons must be zero; otherwise the sum of the 
spin vectors of all neutrinos in the lattice plus the spin vector of the 
electric charge of a $\mu^\pm$ meson could not be $\hbar$/2, as it must be.

\begin{figure}[h]
\vspace{0.4cm}
\hspace{.7cm}
\includegraphics{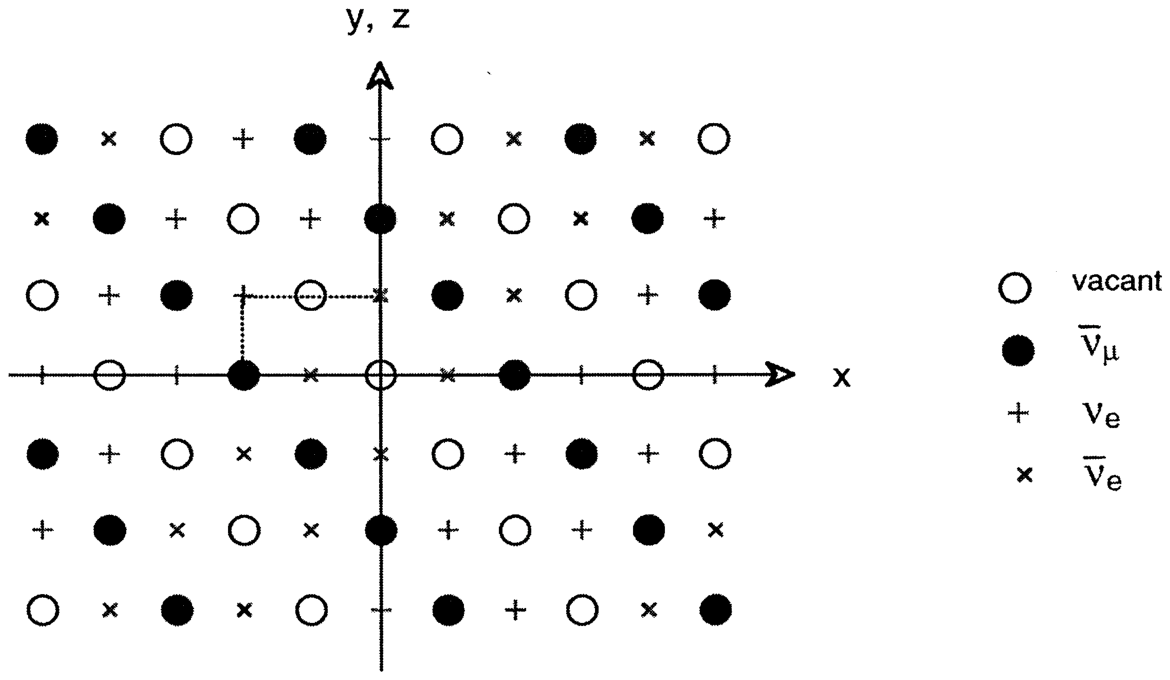}
\vspace{-.2cm}
\begin{quote}
Fig.\,1. The $\mu^+$ meson lattice. The dotted rectangle marks the 
front-side of the  $\pi^\pm$ meson lattice in Fig.\,5 of [2].
\end{quote}
\end{figure}

   In order to show that the angular momentum vectors around a central 
axis of the lattice caused by the spin vectors of all neutrinos in the 
$\mu$ meson lattice is zero we have to consider this lattice in detail, 
Fig.\,1. As we have learned in [3] the, say, $\mu^+$ meson lattice is 
obtained from 
the cubic neutrino lattice of the $\pi^+$ meson through the removal of 
all $\nu_\mu$ neutrinos. That means that in the $\mu^+$ meson lattice are then N/4 = (2.854$\cdot 
10^9$)/4 vacancies at the location where originally  
$\nu_\mu$ neutrinos were. Each vacancy is surrounded in the xy plane by 
combinations of four electron neutrinos. In the z-direction there is on 
top as well as below each vacancy another electron neutrino. The same 
applies for the electron neutrinos around each antimuon neutrino, which 
are the centers of the cells of a $\mu^+$ meson lattice. The cells of 
the $\mu$ meson lattice are octahedrons, two pyramids joint at their 
square base.  As can be seen on Fig.\,1 
each lattice point in the upper right quadrant has at the opposite 
position in the lower left quadrant the \emph{same} type of neutrino. The 
same applies in the upper left and lower right quadrant.

   As is well-known it appears that only left-handed neutrinos and 
right-handed antineutrinos exist, at least as long as the neutrinos are 
massless. Suppose this also holds in the case of neutrinos with a rest 
mass. The 
angular momentum vectors originating from the spin of the neutrinos would 
then 
not cancel on a diagonal through the center of the lattice. The spin 
vectors on either side of the same length on the diagonal are then from 
the same type of neutrino (Fig.\,1) and therefore point in the same 
direction and do not cancel.
   However, the polarization vector of the spin of a neutrino depends on 
the direction of the velocity of the neutrino, because the helicity H is 
given by

\begin{equation} H = \vec{P}\cdot\vec{v}/Pv ,\end{equation}
where $\vec{P}$ is the polarization vector. If only left-handed neutrinos 
(H = $\mathrm{-}$$\beta$) and right-handed antineutrinos (H = $+\beta$) 
exist, the 
direction of $\vec{P}$ must be reversed if the direction of motion of the 
neutrinos during their oscillation in the $\mu$ meson lattice is, in the 
lower left quadrant of Fig.\,1, opposite to the direction of motion in the 
upper right quadrant. This change of the direction of motion follows from 
the equation of motion for the displacements $u_n$ of the lattice points
in Eq.(7) of [2]

\begin{equation}u_n = Ae^{i(\omega t\,\, +\,\, n\phi)} ,\end{equation}
 from which follows that $\dot{u}_n = v_n = i\omega u_n$. The frequencies are given by 

\begin{equation} \nu_n = \nu_0 \phi_n ,\end{equation}
as in Eq.(19) of [2]. Since n$\phi$ = kx,  $\phi_n$ is proportional to x 
with x = n\emph{a}, where \emph{a} is the lattice constant. It follows 
that the direction of motion of the neutrinos 
in the upper right quadrant ($\phi> 0$) is opposite to the direction of 
motion of the neutrinos in the lower left ($\phi < 0$) quadrant. 
Consequently the angular 
momentum vectors around the center of the lattice caused by the spin of 
the neutrinos in the $\mu$ meson lattice are opposite and of the same 
magnitude, they cancel. The only point without an opposite is the point at 
the center of the lattice, at which there is no neutrino, so the center 
does not contribute a spin vector. The sum of the angular momentum vectors 
caused by 
the spin of all neutrinos in the $\mu$ meson lattice is zero, $\sum_{i} 
j(n_i)$ = 0. Together with $\sum_{i} j(\nu_i)$ = 0, as shown above, we 
arrive from Eq.(1) at 

\begin{equation}  j(\mu^\pm) = j(e^\pm) .\end{equation}
The spin s = 1/2 of the $\mu^\pm$ mesons is caused exclusively by the spin 
of the electric charge that a $\mu$ meson carries. Crucial for this point 
is the absence of a neutrino at the center of the $\mu$ meson lattice.

   The cubic lattice of the $\pi^\pm$ mesons is easily recovered from the 
$\mu$ meson lattice by filling all vacancies with either $\nu_\mu$ (or 
$\bar{\nu}_\mu$) neutrinos. The 
$\pi^\pm$ mesons do not have spin. As in the $\mu$ meson 
lattice all neutrino spin vectors around the 
center of the cubic $\pi^\pm$ lattice cancel, but for the angular momentum 
$\hbar$/2 of the center neutrino. 
This angular momentum is canceled by the spin of the electric charge which 
the $\pi^\pm$ mesons carry, so s($\pi^\pm$) = 0, as it must be. This 
explanation supercedes the explanation given in [1] for the spin of the 
particles of the neutrino branch which applies only for a static lattice.  

\section{Conclusions}

If the $\mu^\pm$ mesons consist of a lattice of $\bar{\nu}_\mu$ (respectively 
$\nu_\mu$), $\nu_e$ and 
$\bar{\nu}_e$ neutrinos, as is suggested by the decay of the $\pi^\pm$ 
mesons, then it can be shown [3] that the theoretical mass of the $\mu^\pm$ mesons is, 
within 1\%, equal to the measured mass of the $\mu^\pm$ mesons. Using the octahedronal lattice 
structure of the $\mu^\pm$ mesons suggested by the determination of the 
mass of the $\mu^\pm$ mesons it follows that the angular momentum vectors 
of all longitudinal lattice oscillations are zero and that the sum of the 
angular momentum vectors caused by the spin of the neutrinos of the 
lattice, taken around the center of the lattice, is also zero. The only 
contribution to the intrinsic angular momentum of the $\mu$ mesons comes 
from the spin of the electric charge which the $\mu^\pm$ mesons carry. 
Consequently the spin of the $\mu$ mesons is s($\mu^\pm$) = 1/2, as it must be. We note that the spin of 
the $\mu$ mesons can be explained, without any additional assumption, 
from  the structure of the 
$\mu$ mesons which we have used for the explanation of the mass of 
the $\mu$ mesons.

    Both the  $\pi^\pm$ mesons and the $\mu^\pm$ mesons carry an electric charge. The $\pi^\pm$ mesons  do not have spin, 
whereas  the   $\mu^\pm$ mesons have spin 1/2. The presence or 
absence of a neutrino at the center of the lattice makes the difference. 
The spin of the electric charge in $\pi^\pm$ is canceled by the spin of 
the central neutrino, whereas the spin of the electric charge in $\mu^\pm$ 
remains because there is no central neutrino to cancel the spin of the electric charge.

\bigskip

\noindent
\textbf{REFERENCES}

\smallskip
[1] E.L.Koschmieder, arXiv: physics/0301060 (2003), Hadr.J. (to appear).

\smallskip
[2] E.L.Koschmieder, arXiv: physics/0211100 (2002), Chaos, Solitons and \\
\indent 
\quad\, Fractals {\bfseries18},1129 (2003).

\smallskip
[3] E.L.Koschmieder, arXiv: physics/0110005 (2001).

\end{document}